\pdfoutput=1
\documentclass[ERTS, 2026]{ERTS} 

\usepackage{graphicx}
\usepackage{amsmath}

\makeatletter
\AddToHook{cmd/maketitle/before}{\let\thetitle\@title}
\makeatother

\usepackage{listings}
\usepackage{multirow}
\usepackage{array}
\usepackage{rotating}
\usepackage{hyperref}
\usepackage{algorithmic}
\usepackage{algorithm}
\usepackage{amsfonts}

\usepackage{cellspace}
\usepackage{xcolor} 

\definecolor{codegreen}{rgb}{0,0.6,0}
\definecolor{codegray}{rgb}{0.5,0.5,0.5}
\definecolor{codepurple}{rgb}{0.58,0,0.82}
\definecolor{backcolour}{rgb}{0.95,0.95,0.92}

\lstdefinestyle{mystyle}{
    backgroundcolor=\color{backcolour},   
    commentstyle=\color{codegreen},
    keywordstyle=\color{magenta},
    numberstyle=\tiny\color{codegray},
    stringstyle=\color{codepurple},
    basicstyle=\footnotesize\ttfamily,
    breakatwhitespace=false,         
    breaklines=true,                 
    captionpos=b,                    
    keepspaces=true,                 
    numbers=left,                    
    numbersep=5pt,                  
    showspaces=false,                
    showstringspaces=false,
    showtabs=false,                  
    tabsize=2
}

\newtheorem{definition}{Definition}
\newtheorem{notation}{Notation}
\newtheorem{example}{Example}
\newtheorem{property}{Property}

\newcommand{\comma}{,\allowbreak}
\newcommand{\INP}{\mathit{INP}}
\newcommand{\WGT}{\mathit{WGT}}
\newcommand{\ACC}{\mathit{ACC}}
\newcommand{\blockSize}{bs}

\newcommand{\memInp}{\mathit{inp\_size}}
\newcommand{\memWgt}{\mathit{wgt\_size}}
\newcommand{\memAcc}{\mathit{acc\_size}}

\newcommand{\matA}{A} 
\newcommand{\matB}{B} 
\newcommand{\matC}{C} 
\newcommand{\matX}{X} 
\newcommand{\matY}{Y} 
\newcommand{\matBlock}[2]{#1_{#2}}
\newcommand{\LoadOp}{\textit{LOAD}}
\newcommand{\GemmOp}{\textit{GEMM}}
\newcommand{\AluOp}[1]{\textit{ALU}_{#1}}
\newcommand{\OpElem}{\textit{op}}
\newcommand{\StoreOp}{\textit{STORE}}
\newcommand{\BGemmOp}{\textit{bGEMM}}
\newcommand{\BAluOp}[1]{\textit{bALU}_{#1}}
\newcommand{\AddOp}{\textit{ADD\_ACC}}
\newcommand{\MatrixToBlock}{\mathit{MatrixToBlockIndex}}

\begin{document}

\title{Compilation and Execution of an Embeddable YOLO-NAS on the VTA}

\author{
    Anthony Faure-Gignoux\authorNumber{1},
    Kevin Delmas\authorNumber{2},
    Adrien Gauffriau\authorNumber{1},
    Claire Pagetti\authorNumber{2}}
\address{
    $^1${Airbus}, $^2${ONERA}
}


\maketitle

\chead{\thetitle}

\pagestyle{fancy}

\thispagestyle{plain}

\licenseFootnote{Anthony Faure-Gignoux~et al}

\begin{abstract}%
    \sloppy
    The Versatile Tensor Accelerator (VTA)
    is a promising open-source FPGA-based accelerator to execute
    Convolutional Neural Networks (CNNs)
    for safety-critical domains.
    The purpose of this paper is to formalise and
    improve our VTA stand-alone compiler to fully automate the compilation of
    large CNNs.
    The proposed extensions are illustrated on a YOLO-NAS object detection model.
\end{abstract}

\section{Introduction}
The integration of Machine Learning (ML) into safety-critical systems is of growing importance for complex tasks like object recognition and speech-to-text.
However, the substantial and ever-increasing computational resources required by ML applications present a significant deployment challenge.
This issue is particularly acute in domains constrained by strict hardware limitations, such as aeronautics.
Even multi-core CPUs, recently introduced in aeronautics, 
often struggle to provide the necessary performance for ML workloads.
Therefore, dedicated hardware accelerators, featuring highly parallel computing resources, are emerging as a viable solution.

\subsection{Versatile Tensor Accelerator}
\sloppy
The Versatile Tensor Accelerator (VTA)~\cite{VTA_paper}, an FPGA-based accelerator, emerges as a promising open-source solution.
This is confirmed by the choice of the aeronautical company Daedalean 
to use a VTA-inspired accelerator for running its Convolutional Neural Networks (CNNs) \cite{daedalean}.
The VTA's design is available in two primary forms: a High-Level Synthesis (HLS) \cite{hls} description for FPGA implementation (e.g., on an UltraScale+) and a CHISEL \cite{chisel} description for cycle-accurate simulation. This ecosystem is complemented by a C++ functional simulator.

Executing Machine Learning (ML) models on the VTA requires a compilation chain to transform them into compatible, executable binaries.
The initial compiler was intertwined with TVM \cite{tvm} and not compatible with the certification approach according to DO-178C \cite{DO178C}.
For this reason, we previously developed a stand-alone VTA compilation chain \cite{standalone_vta} 
that converts NN operations into data and instruction binaries executable by the VTA.

\subsection{Stand-alone VTA compiler overview}
The VTA is used to execute the \emph{inference} of
offline trained neural networks.
This means that the CNN structure, the layers, the shape of tensors exchanged between layers, the weights and the activation functions
are fixed.
In addition, the VTA can only execute \emph{quantized} models and, in the configuration we have selected, the inputs, weights and biases are in int32.
Because everything is known a priori, except the input,
the compilation is in charge of computing 1) the instructions,
2) the data organisation, mapping and transfer between
the DRAM and the on-chip accelerator memory,
3) the ordering of instructions launched on the VTA by the CPU and
possibly some CPU processing between successive VTA instructions (e.g. to reshape data).

\begin{figure}[htbp] 
    \centering
    \includegraphics[width=.8\linewidth]{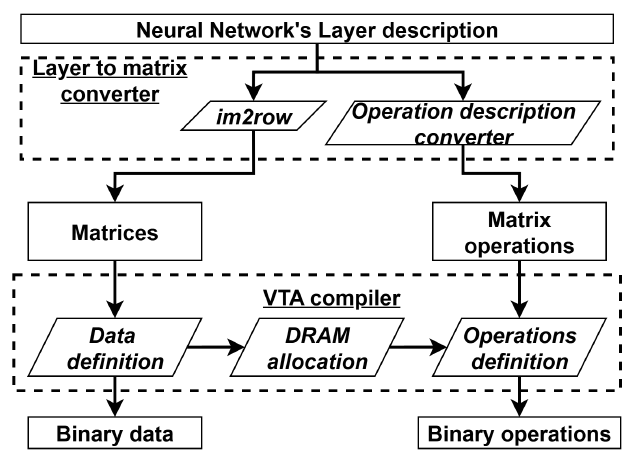}
    \caption{Current VTA compilation chain}
    \label{fig:vta_full_compilation_chain}
\end{figure}

Prior to writing this paper, our stand-alone VTA compiler was only capable of compiling single layers whose weights and inputs fit within the on-chip memory.
Practically, as illustrated in Figure \ref{fig:vta_full_compilation_chain},
each  NN layer (e.g., a convolution followed by an activation)
is converted
into matrix operations (e.g., matrix multiplication and element-wise operations)
with the well-known mathematical transformation $\mathit{im2row}$ \cite{im2row}.

Then,
the VTA compiler converts those matrix operations into instructions.
In effect, 
the matrices are translated into static vectors with fixed size and arranged 
in the precise order needed for computation. 
Those vectors are mapped within the global DRAM according to a pre-defined algorithm. Finally, matrix operations are compiled 
into binary VTA instructions. 

\subsection{Formalising and extending the stand-alone VTA toolchain}
To present the stand-alone VTA compilation chain \cite{standalone_vta}, we explained 1) how data and instructions were allocated to the DRAM;
2) how VTA instructions were encoded with the VTA syntax (e.g., manipulating instruction, UOP),
3) how such instructions were executed on the hardware and their effect on
the vectors stored in the on-chip memory.
Such description was precise, formal and accurate, but rather low level.

The first contribution of the paper extends the formalisation and makes the link between the mathematical operations, their decomposition into blocks, the atomic operations made by the VTA down to a VTA intermediate representation (which corresponds to the description level of our former work  \cite{standalone_vta}).
This can be seen as the compilation passes.

\begin{figure}[htb] 
    \centering
    \includegraphics[width=.5\linewidth]{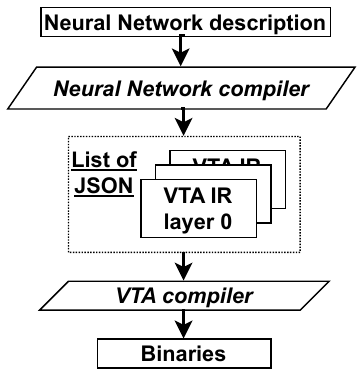}
    \caption{Extended VTA compilation chain}
    \label{fig:extended_vta_compilation_chain}
\end{figure}

The second contribution of this work is the extension of the toolchain to first allow the compilation of larger layers, in the sense that the matrices
exceed the on-chip buffer capacities and requires a sequence of VTA offloads. 
The second extension concerns the execution of complete CNNs.
We managed to compile and execute the LeNet-5 \cite{lenet} but with manual intervention to reshape data between layers.
The objective here is to
automate the inter-layer processing.
The experiments are done on YOLO-NAS~\cite{yolo_all}.
All these extensions are open-source~\cite{onera_github}.

Section \ref{sec:background} recalls how the VTA works but in a more formal way compared to previous work.
Section \ref{sec:formal_op} defines the VTA atomic operations
and how they implement block matrix operations.
Section \ref{sec:ir_ebnf} presents the VTA Intermediate Representation (VTA IR).
Section \ref{sec:extension} details the extended automated compilation toolchain while  Section \ref{sec:strategy} briefly outlines  matrix partitioning strategies required for large matrices.
Finally, Section \ref{sec:experiments}
evaluates the compilation chain on a YOLO-NAS.

\section{Background}
\label{sec:background}
This section aims at presenting an overview of the VTA and reminding matrix decomposition into blocks.

\subsection{VTA architecture and operations}
\begin{figure}[htb] 
    \centering
    \includegraphics[width=.7\linewidth]{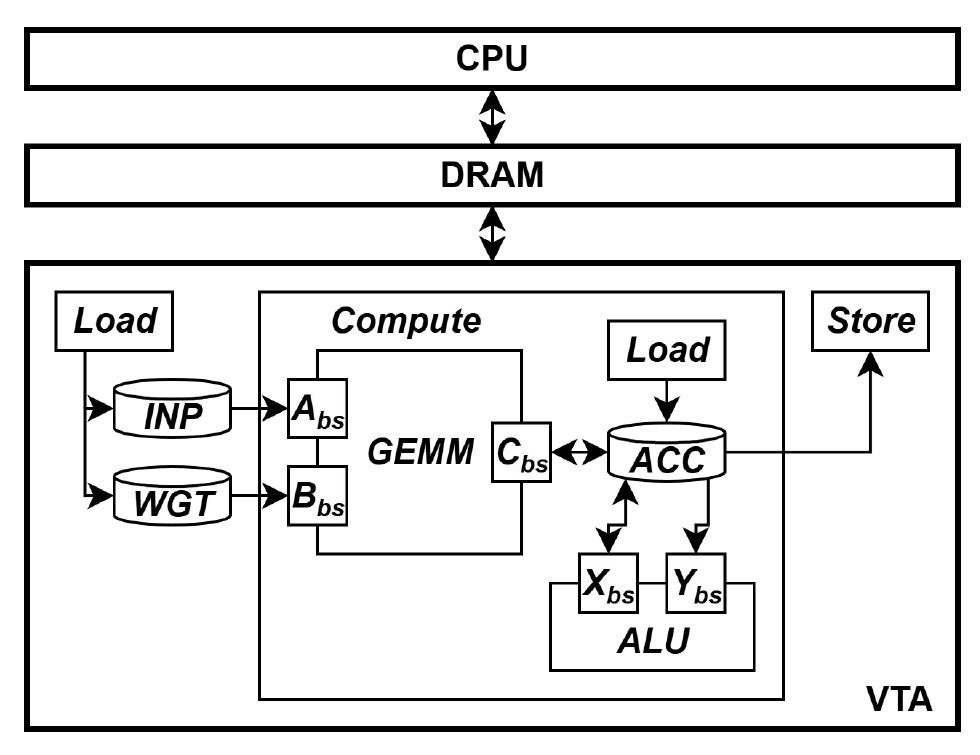}
    \caption{The VTA architecture}
    \label{fig:vta_highlevel_architecture}
\end{figure}

The VTA, shown in Figure \ref{fig:vta_highlevel_architecture}, is a co-processor that works concurrently with the CPU to accelerate matrix computations. 
Both the VTA and the CPU share access to a global DRAM memory.

The VTA comprises an on-chip SRAM memory and three main modules: a \emph{Load} module to transfer data from DRAM into local SRAM, a \emph{Compute} module to perform operations, and a \emph{Store} module to write results from SRAM back to DRAM.

\subsubsection{Data manipulation}
For the sake of clarity, we consider that the VTA operates on two fixed-size data types: $\blockSize \times \blockSize$ matrices and $1 \times \blockSize$ vectors. 
This \emph{block size} ($\blockSize$) is imposed by the VTA hardware configuration.
The on-chip SRAM is decomposed into three buffers. 
$\INP$ (input) contains up to $\memInp$\footnote{$\memInp = 2^{\mathit{LOG\_INP\_BUFF\_SIZE}}/( (\blockSize \times \blockSize) \times 2^{\mathit{LOG\_INP\_WIDTH}})$}
\emph{int32} matrices
(i.e., $\emph{int32}^{\blockSize} \times \emph{int32}^{\blockSize}$).
Similarly, $\WGT$ (weight) contains up to $\memWgt$ int32 matrices,
and $\ACC$ (accumulator) contains up to $\memAcc$ int32 vectors (i.e.,  $\emph{int32}^{\blockSize}$).
\begin{definition}[Organization of data]
  Let $R$ be either the $\INP$, $\WGT$, or $\ACC$ buffer
    of size $r$ (e.g., $r = \memInp$ if $R = \INP$).
    $\forall i \in [\![ 0, r -1 ]\!]$, $R@i$ represents the $i$-th data
    stored in $R$.
    $R@i$ is a $\emph{int32}^{\blockSize}$ vector if $R = \ACC$
    or a $\emph{int32}^{\blockSize} \times \emph{int32}^{\blockSize}$
    matrix otherwise. Moreover, $R(i)$ refers to the value of the i-th vector.
    Similarly, within the DRAM, we consider contiguous memory areas,
    often denoted $Z$ with a size $z$, storing matrices or vectors. $Z@i$ is the address of the i-th data
    and $Z(i)$ the value.
\end{definition}

$\LoadOp$ copies data from the DRAM to the SRAM.
\begin{definition}[VTA's $\LoadOp$ operator]\label{def:vta_init}
    \sloppy
   Let $R$ be either the $\INP$, $\WGT$, or $\ACC$ buffer
    of size $r$.
    Let $Z$ be a DRAM area of size $z$, 
    let $S = (s_j)_{j < k}$
    be an ordered sequence of $k$ distinct integers, 
    with $k \in [\![ 1, z ]\!]$ and $s_j \in [\![ 0, z -1 ]\!]$.
    $\LoadOp$ loads a subset of data of $Z$ in the buffer $R$
    according to the ordered / filtering given by $S$. Formally:
    $
        \LoadOp \big(R, (Z(s_j))_{j < k} \big) 
        \iff 
        \forall j \in [\![ 0, k -1 ]\!], R@j \gets Z(s_j)
    $.
\end{definition}

\begin{figure}[htb] 
    \centering
    \includegraphics[width=0.6\linewidth]{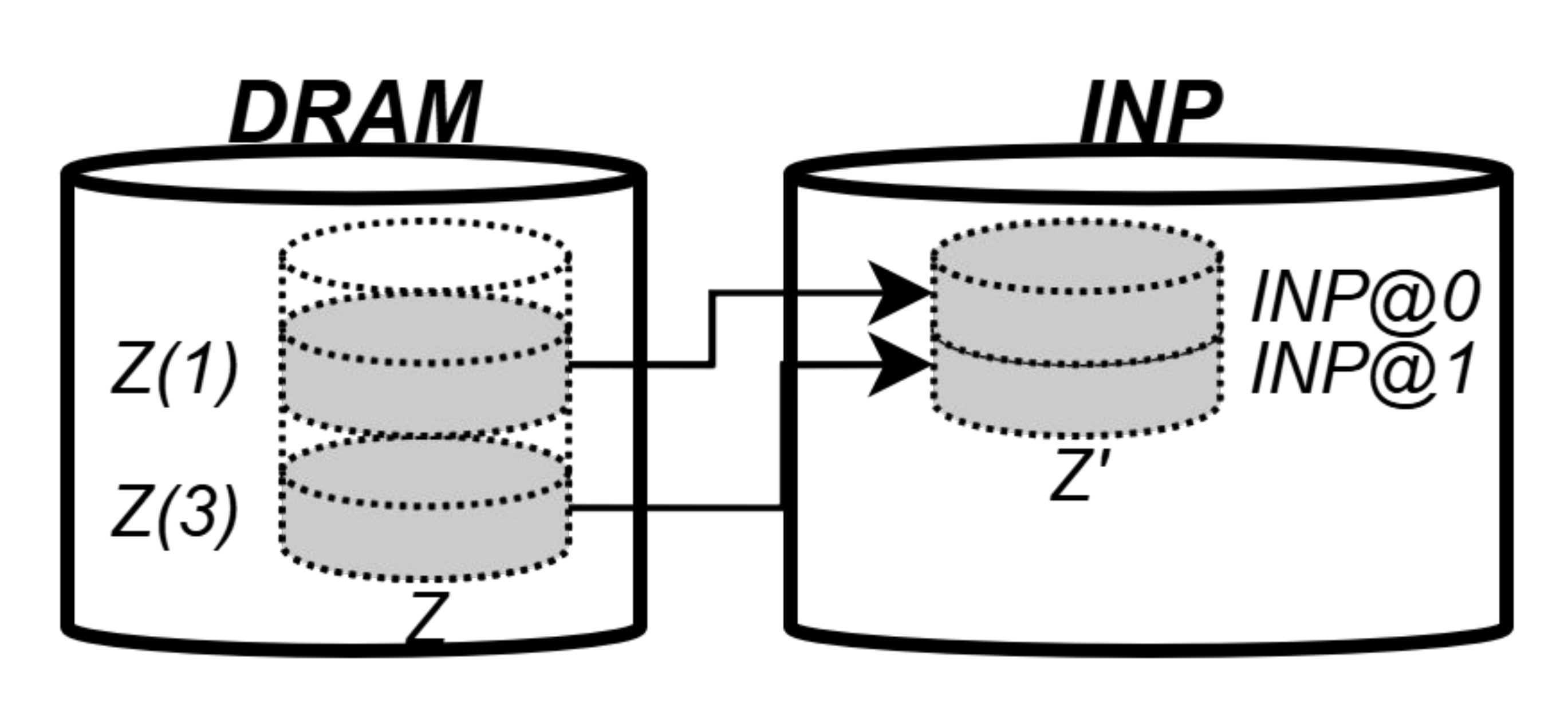}
    \caption{VTA  $\LoadOp$ operator}
    \label{fig:vta_init}
\end{figure}

\begin{example}
  \label{ex:vta_init}
    \sloppy
    Let $Z$ = $\big( Z(i) \big)_{i < 4}$ be a sequence of $\blockSize \times \blockSize$ matrices.
    $\LoadOp \big(\INP, (Z(1), Z(3)) \big)$ results in $\INP@0 \gets Z(1)$ and $\INP@1 \gets Z(3)$ as illustrated in Figure \ref{fig:vta_init}.
\end{example}

The $\StoreOp$ operator writes data stored in
$\ACC$ to the DRAM.
\begin{definition}[VTA's $\StoreOp$ operator]\label{def:vta_store}
    \sloppy
    Let $S = (s_j)_{j < k}$ be an ordered sequence of $k$ distinct integers
    within $[\![ 0, \memAcc -1 ]\!]$.
    %
    Let $Z$ be a DRAM contiguous space storing $k$ int32 vectors.
    %
   Then:
    $
        \StoreOp \big( (\ACC(s_j))_{j < k} \big)
        \iff 
        \forall j \in [\![ 0, k -1 ]\!], Z@j \gets \ACC(s_j)
    $.
\end{definition}
Note that a single operator, $\LoadOp$ or $\StoreOp$, may correspond to multiple VTA instructions.

\subsubsection{Computation operations}
The Compute module comprises three submodules: \emph{Load}, \emph{GEMM} (General Matrix Multiplication), and \emph{ALU} (Arithmetic Logic Unit). 
The GEMM submodule performs matrix-matrix multiplication and accumulation.
A GEMM instruction
provides the buffer addresses for three $\blockSize \times \blockSize$ matrices.
For the sake of clarity, we represent this scheme with
three interfaces ($\matBlock{\matA}{\blockSize}$, $\matBlock{\matB}{\blockSize}$, and $\matBlock{\matC}{\blockSize}$). 
\begin{definition}[VTA's $\GemmOp$ operation]\label{def:vta_gemm}
    Let $\matBlock{\matA}{\blockSize}$, $\matBlock{\matB}{\blockSize}$, $\matBlock{\matC}{\blockSize}$ be $\blockSize \times \blockSize$ matrices.
    $ 
        \GemmOp(\matBlock{\matC}{\blockSize}, \matBlock{\matA}{\blockSize}, \matBlock{\matB}{\blockSize}) 
        \iff 
        \matBlock{\matC}{\blockSize} := \matBlock{\matC}{\blockSize} + \matBlock{\matA}{\blockSize} \times \matBlock{\matB}{\blockSize} 
    $.
\end{definition}

The ALU submodule performs element-wise operations
on two vectors represented with an interface
$\matBlock{\matX}{\blockSize}$ and $\matBlock{\matY}{\blockSize}$.
Five operators are allowed: \emph{MAX} (maximum), \emph{MIN} (minimum), \emph{ADD} (addition), \emph{MUL} (multiplication), and \emph{SHR} (shift right).

\begin{definition}[VTA's $\AluOp{\OpElem}$ operation]\label{def:vta_alu}
    \sloppy
    Let $\matBlock{\matX}{\blockSize}$ and  $\matBlock{\matY}{\blockSize}$
    be 2 vectors.
    The $\AluOp{\OpElem}$ is an element-wise operation applied
    on the two vectors or on  $\matBlock{\matX}{\blockSize}$ and a scalar value $c$.
    We have
    $\begin{cases} 
        \AluOp{\OpElem}( \matBlock{\matX}{\blockSize},  \matBlock{\matY}{\blockSize}) &
        \iff 
         \matBlock{\matX}{\blockSize} := \OpElem (  \matBlock{\matX}{\blockSize},  \matBlock{\matY}{\blockSize}) \\
        \AluOp{\OpElem}( \matBlock{\matX}{\blockSize}, c) 
       & \iff 
         \matBlock{\matX}{\blockSize} := \OpElem ( \matBlock{\matX}{\blockSize}, c ) 
    \end{cases}$
\end{definition}

\begin{example}
  \label{ex:vta_alu}
    \sloppy
    We set $\blockSize = 2$ and
    we perform first $\AluOp{\mathit{max}}(X_0, X_1)$
    then $\AluOp{\mathit{max}}( X_1, 3)$.
    The updated values are in bold.
    $$ 
    \begin{cases}
       X_0= [0 & 8] \\
       X_1 = [2 & 1]
    \end{cases}
    \xrightarrow[]{}
    \begin{cases}
        [\textbf{2}& \textbf{8}] \\
        [2 &1]
    \end{cases}
    \xrightarrow[]{}
    \begin{cases}
        [2 & 8] \\
        [\textbf{3} & \textbf{3}]
    \end{cases}
    $$
\end{example}

\subsection{Background on block matrix}
The VTA loads matrices larger than $\blockSize \times \blockSize$. 
To perform operations on those matrices, they are partitioned into block matrices on which VTA operations are performed.
Let us introduce notations on block matrices.

\begin{definition}[Block matrix]\label{def:block_matrix}
    Let $\matA = [\matA(i,j)]$ be a $m \times n$ matrix such that $\blockSize | m$ and $\blockSize | n$.
    $\matA$ can be seen as a $\alpha \times \beta$-block matrix with $\alpha = m / \blockSize$ and $\beta = n / \blockSize$ such that:
    \begin{equation*}
        \matA =
        \begin{bmatrix}
            \matBlock{\matA}{0} & \hdots & \matBlock{\matA}{\beta - 1} \\
            \hdots & \matBlock{\matA}{i \times \beta + j} & \hdots \\
            \matBlock{\matA}{(\alpha-1) \times \beta} & \hdots & \matBlock{\matA}{\alpha \times \beta - 1}
        \end{bmatrix}
    \end{equation*}
    Where for $u = i \times \blockSize$ and $v = j \times \blockSize$,
 $\matBlock{\matA}{i \times \beta + j} =$
    \begin{equation*}
        \begin{bmatrix}
            \matA(u, v) & \hdots & \matA(u, v + \blockSize - 1) \\
            \hdots & \hdots & \hdots \\
            \matA(u + \blockSize - 1, v) & \hdots & \matA(u + \blockSize - 1, v + \blockSize - 1)
        \end{bmatrix}
    \end{equation*}
   \end{definition}

\begin{example}[Block matrix]\label{ex:block_matrix}
  Let $\blockSize = 2$ and transform a $2 \times 4$
  matrix $\matA$ into $1 \times 2$-block matrix:
    \begin{equation*}
        \matA =
        \begin{bmatrix}
            \matA(0, 0) & \matA(0, 1) & \matA(0, 2) & \matA(0, 3) \\
            \matA(1, 0) & \matA(1, 1) & \matA(1, 2) & \matA(1, 3)
        \end{bmatrix}
        =
        \begin{bmatrix}
            \matBlock{\matA}{0} & \matBlock{\matA}{1}
        \end{bmatrix}
    \end{equation*}
    With:
    \begin{equation*}
        \matBlock{\matA}{0} =
        \begin{bmatrix}
            \matA(0, 0) & \matA(0, 1) \\
            \matA(1, 0) & \matA(1, 1)
        \end{bmatrix},
        \matBlock{\matA}{1} =
        \begin{bmatrix}
            \matA(0, 2) & \matA(0, 3) \\
            \matA(1, 2) & \matA(1, 3)
        \end{bmatrix}
    \end{equation*}
\end{example}

$\MatrixToBlock$
relates an element in a matrix
to the corresponding element in its block decomposition.
\begin{definition}[Matrix to block element relation]\label{def:matrixToBlock}
    Let $\matA$ be a $m \times n$ matrix such that $\blockSize | m$ and $\blockSize | n$.
    %
    $\forall (i,j) \in [\![0, m-1]\!] \times [\![0, n-1]\!]$,
    the function $\MatrixToBlock$ maps the indices of the element $\matA(i,j)$ to those in the block $\matBlock{\matA}{k}(u,v)$, with $k \in [\![0, \alpha \times \beta - 1]\!], (u,v) \in [\![0, \blockSize-1]\!] \times [\![0, \blockSize-1]\!]$:
    $$(i,j) \mapsto (k, (u,v))$$
    $$(i,j) \mapsto \Big ( \Big \lfloor \frac{i}{\blockSize} \Big \rfloor \times \beta + \Big \lfloor \frac{j}{\blockSize} \Big \rfloor, (i \mod \blockSize, j \mod \blockSize) \Big )$$
\end{definition}

\begin{example}[Matrix to block element relation]\label{ex:matrixToBlock}
    \sloppy
    Let us continue Example~\ref{ex:block_matrix} with $\blockSize = 2$ and $\matA$.
    Applying $\MatrixToBlock(1,2) =$
    $\Big ( 
        \Big \lfloor \frac{1}{2} \Big \rfloor \times \beta + \Big \lfloor \frac{2}{2} \Big \rfloor 
        \comma 
        (1 \mod 2, 2 \mod 2) 
    \Big ) = (1, (1,0)) 
    $,
    we find that $\matA(1,2) = \matBlock{\matA}{1}(1,0)$.
\end{example}

Finally,  $\matX(i)$ is used to access the $i$-th row of a matrix $\matX$.
\begin{notation}[Vectors and elements of a matrix]\label{not:matrix_elem}
    Let $\matX$ be a $m \times n$ matrix with $\forall (i,j) \in [\![ 0, m -1 ]\!] \times [\![ 0, n -1 ]\!], \matX(i,j)$ be its elements.
    The vector created from the $i$-th row of $\matX$ is:
    $\forall i \in [\![0, m-1]\!], \matX(i) = (\matX(i,j))_{j < n}$.
\end{notation}

\section{Formalisation of atomic operations} \label{sec:formal_op}
Let us formalise how matrix-level operations are transformed into a set of VTA operations.

\subsection{GEMM for matrices}
In this section, we consider
$\matA$, $\matB$, $\matC$ matrices loaded in $\INP$, $\WGT$ and $\ACC$ respectively with
$\matA$ an $\alpha \times \lambda$-block matrix,
$\matB$ a $\lambda \times \beta$-block matrix
and $\matC$ an $\alpha \times \beta$-block matrix.
The computation $\matC := \matC + \matA \times \matB$ ($\iff \BGemmOp(\matC, \matA, \matB)$) is performed by applying several VTA's $\GemmOp$ operations (see Definition \ref{def:vta_gemm}).
Moreover, those operators are \emph{independent} as
their order of execution does not have any impact on the result.

\begin{definition}[$\BGemmOp$ operation]\label{def:block_matmul}
    \sloppy
  $\BGemmOp(\matC, \matA, \matB)$
    is the set of operations:
    $ 
        \{ \GemmOp(\matBlock{\matC}{i \times \beta + j}, \matBlock{\matA}{i \times \lambda + k}, \matBlock{\matB}{k \times \beta + j}) \allowbreak | i \in [\![0, \alpha -1]\!], j < [\![0, \beta -1]\!], k \in [\![0, \lambda -1]\!]
        \}
    $.
\end{definition}


%

\begin{example}\label{ex:block_matmul}
    \sloppy
    Let us continue Example \ref{ex:block_matrix} with $\blockSize = 2$.
    $\matA$ and $\matC$ are $1 \times 2$-block matrices,
    $\matB$ is a $2 \times 2$-block matrix: 
    \begin{equation*}
        \matA =
        \begin{bmatrix}
            \matBlock{\matA}{0} & \matBlock{\matA}{1}
        \end{bmatrix}
        ,
        \matB =
        \begin{bmatrix}
            \matBlock{\matB}{0} & \matBlock{\matB}{1}\\
            \matBlock{\matB}{2} & \matBlock{\matB}{3}
        \end{bmatrix}
        ,
        \matC =
        \begin{bmatrix}
            \matBlock{\matC}{0} & \matBlock{\matC}{1}
        \end{bmatrix}
    \end{equation*}
     $\BGemmOp(\matC, \matA, \matB)$ consists in computing: 
     $$\begin{bmatrix}
         \matBlock{\matC}{0} + \matBlock{\matA}{0} \times \matBlock{\matB}{0} + \matBlock{\matA}{1} \times \matBlock{\matB}{2} & \matBlock{\matC}{1} + \matBlock{\matA}{0} \times \matBlock{\matB}{1} + \matBlock{\matA}{1} \times \matBlock{\matB}{3}
     \end{bmatrix}$$
     which is done with:
    $\{ \GemmOp(\matBlock{\matC}{0}, \matBlock{\matA}{0},\matBlock{\matB}{0}),\GemmOp(\matBlock{\matC}{0}, \matBlock{\matA}{1},$ $\matBlock{\matB}{2}) \comma \GemmOp(\matBlock{\matC}{1}, \matBlock{\matA}{0}, \matBlock{\matB}{1})\comma  \GemmOp(\matBlock{\matC}{1}, \matBlock{\matA}{1}, \matBlock{\matB}{3}) \}$.
\end{example}

In case of multiplication with a scalar (see Definition~\ref{def:block_matmul_scalar}), the scalar has to be converted into a block, because the VTA hardware can only perform matrix-matrix multiplications.

\begin{definition}[$\BGemmOp$ with scalar]\label{def:block_matmul_scalar}
    \sloppy
    Let $b$ be a scalar. We define the diagonal matrix
    $B = b \times I_{\blockSize}$ with $I_{\blockSize}$ the identity matrix of size $\blockSize$.
    The operation     $
        \BGemmOp(\matC, \matA, b) \iff \matC := \matC + \matA \times b
    $,
    is the set of operations:
    $ 
        \{ \GemmOp(\matBlock{\matC}{i \times \beta + j}, \matBlock{\matA}{i \times \lambda + k}, B) \allowbreak | i \in [\![0, \alpha -1]\!], j < [\![0, \beta -1]\!], k \in [\![0, \lambda -1]\!]
        \}
    $.
\end{definition}

\begin{example}
  \label{ex:block_matmul_scalar}
    \sloppy
    Taking back Example~\ref{ex:block_matrix} with $\blockSize = 2$.
    This time $B =\begin{bmatrix}
            b & 0 \\
            0 & b
        \end{bmatrix}$.
    The operation $\BGemmOp(\matC, \matA, b)$ results in:
    $$\begin{bmatrix}
         \matBlock{\matC}{0} + \matBlock{\matA}{0} \times B + \matBlock{\matA}{1} \times B & \matBlock{\matC}{1} + \matBlock{\matA}{0} \times B + \matBlock{\matA}{1} \times B
    \end{bmatrix}$$
    which is done with:
    $ \{ \GemmOp(\matBlock{\matC}{0}, \matBlock{\matA}{0}, \matB) \comma \GemmOp(\matBlock{\matC}{0}, \matBlock{\matA}{1},$ $ \matB) \comma \GemmOp(\matBlock{\matC}{1}, \matBlock{\matA}{0}, \matB) \comma \GemmOp(\matBlock{\matC}{1}, \matBlock{\matA}{1}, \matB) \}$.
\end{example}

\subsection{ALU for vectors and matrices}
Applying element-wise operations on 
vectors of size larger than $\blockSize$ loaded in  $\ACC$
consists in executing independent VTA-ALU operations.

\begin{definition}[$\BAluOp{\OpElem}$ operation]
  \label{def:alu_vector-vector}
    \sloppy
     Let $\matX$ and $\matY$
be 2 vectors of size $\beta\times \blockSize$
loaded in  $\ACC$.
    $
\BAluOp{\OpElem}(\matX, \matY) \iff 
        X := \OpElem ( X, Y)$
consists of 
$
        \{ 
            \AluOp{\OpElem}(\matBlock{\matX}{i \times \beta}, \matBlock{\matY}{i \times \beta})
            | i \in [\![ 0, \beta -1 ]\!] 
        \}
    $
    Similarly, for an integer $c$, 
    $
        \BAluOp{\OpElem}(\matX, c) = 
        \{ 
            \AluOp{\OpElem}(\matBlock{\matX}{i \times \beta}, c)
            | i \in [\![ 0, \beta -1 ]\!] 
        \}
    $.
\end{definition}

\begin{example}
  \label{ex:alu_vector-vector}
    \sloppy
    Let
    $\matX =
    \begin{bmatrix}
        \matBlock{\matX}{0} & \matBlock{\matX}{1}
    \end{bmatrix}$ and $\matY =
    \begin{bmatrix}
        \matBlock{\matY}{0} & \matBlock{\matY}{1}
    \end{bmatrix}$.
     Computing $X := \textit{max} ( X, Y)$
is done with
     $
        \{
            \AluOp{\mathit{max}}( \matBlock{\matX}{0}, \matBlock{\matY}{0} ) \comma 
            \AluOp{\mathit{max}}( \matBlock{\matX}{1}, \matBlock{\matY}{1} )
        \} 
    $.
\end{example}

To simplify the description of matrix additions, we propose a kind of \emph{syntactic sugar} in the form of a new operation named  $\AddOp$.
\begin{definition}[$\AddOp$ operation]\label{def:vta_add}
    \sloppy
    Let $\matX$ and $\matY$ be $\alpha \times \beta$-block matrices.
    The operation $\matX := \matX + \matY$ $\iff \AddOp(\matX, \matY)$
    is performed by the set of operations:
    $$
        \{ \BAluOp{\mathit{add}}(\matX(i), \matY(i)) | i \in [\![ 0, \alpha -1 ]\!] \}
    $$
\end{definition}


\begin{example}
  \label{ex:add_2_matrices}
    \sloppy
  Let us apply the operation  $\AddOp(X, Y)$:
    \begin{equation*}
        X =
        \begin{bmatrix}
            \matBlock{X}{0} & \matBlock{X}{1} \\
            \matBlock{X}{2} & \matBlock{X}{3}
        \end{bmatrix}
        +
        \begin{bmatrix}
            \matBlock{Y}{0} & \matBlock{Y}{1} \\
            \matBlock{Y}{2} & \matBlock{Y}{3}
        \end{bmatrix}
        =
        \begin{bmatrix}
            \matBlock{X}{0} + \matBlock{Y}{0} & \matBlock{X}{1} + \matBlock{Y}{1} \\
            \matBlock{X}{2} + \matBlock{Y}{2} & \matBlock{X}{3} + \matBlock{Y}{3} 
        \end{bmatrix}
    \end{equation*}
\end{example}

\section{VTA Intermediate Representation} \label{sec:ir_ebnf}
Matrix operations associated to a NN layer
are declared within a JSON file
used by the stand-alone VTA compiler \cite{standalone_vta} 
to produce the VTA operations. 
Such JSON file can be seen as a \emph{VTA Intermediate Representation} (VTA IR). 
The syntax of the VTA IR is described using EBNF (Extended Backus-Naur Form). 
Listing~\ref{lst:ebnfJSONprim} begins by defining the primitive types used within this format.

\begin{lstlisting}[caption={Primitive types (EBNF for JSON input)}, label={lst:ebnfJSONprim}]
id ::= [a-zA-Z_], [a-zA-Z0-9_]*;
string ::= '(*@\textbf{"}@*)', id, '(*@\textbf{"}@*)';
integer ::= (-)?, [0-9]+;
int_pair ::= '(*@\textbf{[}@*)', integer, '(*@\textbf{,}@*)', integer, '(*@\textbf{]}@*)';
hex ::= '(*@\textbf{"}@*)', [0-9a-fA-F]+, '(*@\textbf{"}@*)';
path ::=  '(*@\textbf{"}@*)', (('(*@\textbf{/}@*)')?, [a-zA-Z0-9_.-]+, '(*@\textbf{/}@*)')*, 
    id, '(*@\textbf{.bin}@*)', '(*@\textbf{"}@*)';
\end{lstlisting}

As detailed in Listing~\ref{lst:ebnfJSONmatrix}, a matrix is specified by a unique name (\lstinline|string|), its dimensions (rows and columns),
and either a file path to its raw elements for the fixed parameters such as the weights
or keyword \lstinline|"input"|
for the variables exchanged between layers.
The \lstinline|"MATRICES"| field specifies between one and three matrices
(\lstinline|matrix_def|), plus the accumulator vectors (designated as a matrix \lstinline|"output"|).

\begin{lstlisting}[caption={Matrix Definition (EBNF)}, label={lst:ebnfJSONmatrix}]
matrix_def ::= string, '(*@\textbf{: }@*)', '(*@\textbf{[}@*)', 
    integer, '(*@\textbf{,}@*)', integer, '(*@\textbf{,}@*)', ('(*@\textbf{"input"}@*)' | path), 
    '(*@\textbf{]}@*)';
matrix_out ::= string, '(*@\textbf{: }@*)', '(*@\textbf{[}@*)', 
    integer, '(*@\textbf{,}@*)', integer, '(*@\textbf{,}@*)', '(*@\textbf{"output"}@*)', '(*@\textbf{]}@*)';
matrices ::= '(*@\textbf{"MATRICES"}@*)', '(*@\textbf{: }@*)', '(*@\textbf{\{}@*)', 
    matrix_def, '(*@\textbf{,}@*)', (matrix_def, '(*@\textbf{,}@*)',
   (matrix_def, '(*@\textbf{,}@*)')?)?, matrix_out, '(*@\textbf{\}}@*)';
\end{lstlisting}

For instance, Listing~\ref{lst:ebnfJSONmatEx} shows the declaration of four $16 \times 16$ matrices: $A$, $B$, $X$ and the output $C$.

\begin{lstlisting}[caption={Example for Matrix Definition}, label={lst:ebnfJSONmatEx}]
"MATRICES": {
    "A": [16,16, "input"],
    "B": [16,16, "../workspace/wgt.bin"],
    "X": [16,16, "../workspace/acc.bin"],
    "C": [16,16, "output"]
}
\end{lstlisting}

Listing \ref{lst:ebnfJSONload} provides the specification for initialising the buffers with $\LoadOp$ operator. 
A specific buffer load is defined by the buffer name, the matrix identifier (\lstinline|string|),
and an optional \lstinline|data_list| field
that specifies which parts of the matrix to load.
When omitted, the entire matrix (\lstinline|string|)
is loaded in the buffer.

\begin{lstlisting}[caption={Load Operation (EBNF)}, label={lst:ebnfJSONload}]
data ::= '(*@\textbf{[}@*)', int_pair, '(*@\textbf{,}@*)', integer, '(*@\textbf{]}@*)';
data_list ::= data, ('(*@\textbf{,}@*)', data)*;
load_inp ::= '(*@\textbf{"INP"}@*)',  '(*@\textbf{: }@*)', '(*@\textbf{[}@*)',
    string, ('(*@\textbf{,}@*)', data_list)?, '(*@\textbf{]}@*)';
load_wgt ::= '(*@\textbf{"WGT"}@*)',  '(*@\textbf{: }@*)', '(*@\textbf{[}@*)',
    string, ('(*@\textbf{,}@*)', data_list)?, '(*@\textbf{]}@*)';
load_acc ::= '(*@\textbf{"ACC"}@*)',  '(*@\textbf{: }@*)', '(*@\textbf{[}@*)',
    string, ('(*@\textbf{,}@*)', (data_list | string))?, '(*@\textbf{]}@*)';
load ::= '(*@\textbf{"LOAD"}@*)', '(*@\textbf{: }@*)', '(*@\textbf{\{}@*)', 
    ((load_inp, '(*@\textbf{,}@*)', load_wgt, ('(*@\textbf{,}@*)', load_acc)?) 
    | load_acc), '(*@\textbf{\}}@*)';
\end{lstlisting}

The pseudo-code to load
in buffer $R$
the part of matrix $Z$ given by a \lstinline|data_list|
is provided in Algorithm \ref{algo:load_pseudocode}. 

\begin{algorithm}[hbt]
\caption{Pseudo-code of Load}
\label{algo:load_pseudocode}
\begin{algorithmic}[1]
\renewcommand{\algorithmicrequire}{\textbf{Input:}} 
\REQUIRE $R$, $Z$, $([[a_{i},b_{i}], c_{i}])_{i < k}$
\FOR {$i \gets 0$ until $k$}
    \FOR {$j \gets 0$ until $c_{i}$}
        \STATE $y = i \times k + j$
        \STATE $x = a_{i} + j \times b_{i}$
        \STATE $R@y \gets Z(x)$
    \ENDFOR 
\ENDFOR 
\end{algorithmic}
\end{algorithm}

The \lstinline|load_acc| can also load two matrices in $\ACC$ if two \lstinline|string| identifiers are given.
\begin{example}
  \label{ex:ebnfJSONloadEx2}
    \sloppy
    Let $Y$, $Z$ be two $2 \times 2$ matrices located in the DRAM.
    Listing \ref{lst:ebnfJSONloadEx2} is the syntax to declare the operation:
    $ \LoadOp \big( \ACC, \big( Y(0), Y(1), Z(0), Z(1) \big) \big) $. 
\end{example}

\begin{lstlisting}[caption={Example for Load Operation (two matrices)}, label={lst:ebnfJSONloadEx2}]
"LOAD" : {"ACC": ["Y", "Z"]}
\end{lstlisting}

\sloppy
Listing \ref{lst:ebnfJSONloadEx} builds on
Listing \ref{lst:ebnfJSONmatEx} to load data in the buffers. 
Lines 2 and 3 load the entire $A$ and $B$ matrices, respectively. 
Line 2 corresponds to $\LoadOp \big( \INP, (A(i))_{i<1} \big)$ (see 
Definition \ref{def:vta_init}).
Line 4, in contrast, loads only specific vectors from $C$, more precisely $\LoadOp \big( \ACC, (C(0), C(1), C(4), C(8)) \big)$.

\begin{lstlisting}[caption={Example for Load Operation}, label={lst:ebnfJSONloadEx}]
"LOAD" : {
    "INP": ["A"],
    "WGT": ["B", [[0, 1], 1]],
    "ACC": ["X", [[0, 1], 2], [[4, 4], 2]]
}
\end{lstlisting}

Listing \ref{lst:ebnfJSONgemm} specifies the declaration of the $\GemmOp$ operator. It can multiply a matrix by either another matrix or a scalar. 

\begin{lstlisting}[caption={GEMM Operation (EBNF)}, label={lst:ebnfJSONgemm}]
gemm ::= '(*@\textbf{"GEMM"}@*)', '(*@\textbf{: }@*)', '(*@\textbf{[}@*)', 
    string, '(*@\textbf{,}@*)', string, '(*@\textbf{,}@*)', (integer | string), 
    '(*@\textbf{]}@*)';
\end{lstlisting}

Listing \ref{lst:ebnfJSONgemmEx} continues Listing \ref{lst:ebnfJSONloadEx}
by declaring $\BGemmOp(C, A, B)$ to compute $C = C + A \times B$.
For $C = X + A \times b$, the listing would be \lstinline|"GEMM": ["C", "A", b]|.

\begin{lstlisting}[caption={Example for GEMM Operation}, label={lst:ebnfJSONgemmEx}]
"GEMM": ["C","A","B"]
\end{lstlisting}

Listing \ref{lst:ebnfJSONalu} specifies the \emph{ALU} operations.
We remind there are five of them: Maximum (\lstinline|MAX|), Minimum (\lstinline|MIN|), Addition (\lstinline|ADD|), Multiplication (\lstinline|MUL|), and Arithmetic Shift Right (\lstinline|SHR|).
They support two modes: vector-vector and vector-scalar. 
In vector-scalar mode, they are termed \emph{immediate} (e.g., \lstinline|MAX_IMM|).

\begin{lstlisting}[caption={List of ALU operations (EBNF)}, label={lst:ebnfJSONaluList}]
op_name ::= '(*@\textbf{"MAX"}@*)'|'(*@\textbf{"MIN"}@*)'|'(*@\textbf{"ADD"}@*)'|'(*@\textbf{"MUL"}@*)'|'(*@\textbf{"SHR"}@*)';
op_imm ::= '(*@\textbf{"MAX\_IMM"}@*)' | '(*@\textbf{"MIN\_IMM"}@*)' | '(*@\textbf{"ADD\_IMM"}@*)' | '(*@\textbf{"MUL\_IMM"}@*)' | '(*@\textbf{"SHR\_IMM"}@*)';
alu_vector_vector ::= '(*@\textbf{[}@*)', op_name, '(*@\textbf{,}@*)', '(*@\textbf{[}@*)', int_pair, '(*@\textbf{,}@*)', int_pair, '(*@\textbf{,}@*)', integer,'(*@\textbf{]}@*)','(*@\textbf{]}@*)';
alu_vector_scalar ::= '(*@\textbf{[}@*)', op_imm, '(*@\textbf{,}@*)', '(*@\textbf{[}@*)', int_pair, '(*@\textbf{,}@*)', integer, '(*@\textbf{,}@*)', integer,'(*@\textbf{]}@*)','(*@\textbf{]}@*)';
alu_op ::= alu_vector_vector | alu_vector_scalar;
alu_list ::= alu_op, ('(*@\textbf{,}@*)', alu_op)*;
\end{lstlisting}

Algorithms \ref{algo:ALUvv_pseudocode} and \ref{algo:ALUvs_pseudocode} provide the pseudo-code of the ALU sequence for both vector-vector and vector-scalar operations.
The first argument specifies the operator to apply, followed by an integer pair defining the destination vector's starting index and its per-iteration increment. 
The next argument is either a second integer pair for the source vector (for vector-vector operations) or a single integer for the scalar (for vector-scalar operations). 
A final integer specifies the total number of iterations to perform.

\begin{algorithm}[hbt]
\caption{Pseudo-code of vector-vector ALU}
\label{algo:ALUvv_pseudocode}
\begin{algorithmic}[1]
\renewcommand{\algorithmicrequire}{\textbf{Input:}} 
\REQUIRE $C$, (["$\OpElem_{i}$", [[$a_{i}$,$b_{i}$], [$c_{i}$,$d_{i}$], $e_{i}$]])$_{i < N}$
\FOR {$i \gets 0$ until $N$}
    \FOR {$j \gets 0$ until $e_{i}$}
        \STATE $x = a_{i} + j \times b_{i}$
        \STATE $y = c_{i} + j \times d_{i}$
        \STATE $\BAluOp{\OpElem_{i}}(C(x), C(y))$
    \ENDFOR
\ENDFOR
\end{algorithmic}
\end{algorithm}

\begin{algorithm}[hbt]
\caption{Pseudo-code of vector-scalar ALU}
\label{algo:ALUvs_pseudocode}
\begin{algorithmic}[1]
\renewcommand{\algorithmicrequire}{\textbf{Input:}} 
\REQUIRE $C$, (["$\OpElem_{i}$", [[$a_{i}$,$b_{i}$], $c_{i}$, $e_{i}$]])$_{i < N}$
\FOR {$i \gets 0$ until $N$}
    \FOR {$j \gets 0$ until $e_{i}$}
        \STATE $x = a_{i} + j \times b_{i}$
        \STATE $\BAluOp{\OpElem_{i}}(C(x), c_{i})$
    \ENDFOR
\ENDFOR
\end{algorithmic}
\end{algorithm}

Listing \ref{lst:ebnfJSONaluListEx} provides an example of a list of ALU operations applied to a $6 \times 2$ matrix (named $C$).
Example \ref{ex:ebnfJSONaluListEx} then illustrates the effect of these operations on that matrix.

\sloppy
Line 1 implements the simple vector-vector operation $\BAluOp{\mathit{max}} ( C(0), C(1) )$, while Line 2 implements the vector-scalar operation $\BAluOp{\mathit{max}} ( C(0), 1 )$. 
Line 3 implements $\big( \BAluOp{\mathit{max}} ( C(0 + i \times 2), C(1 + i \times 2) ) \big)_{i < 3}$, and Line 4 implements $\big( \BAluOp{\mathit{max}} ( C(0 + i), 0 ) \big)_{i < 6}$. 
Line 4 is a special case, as it is equivalent to a ReLU (Rectified Linear Unit) operation.

\begin{lstlisting}[caption={Example for a List of ALU operations}, label={lst:ebnfJSONaluListEx}]
["MAX", [[0,0], [1,0], 1]],
["MAX_IMM", [[0,0], 1, 1]],
["MAX", [[0,2], [1,2], 3]],
["MAX_IMM", [[0,1], 0, 6]]
\end{lstlisting}

\begin{example}[ALU operations] \label{ex:ebnfJSONaluListEx}
    Let $C$ be the matrix used by Listing \ref{lst:ebnfJSONaluListEx} and $Li$ be the operation declared on the $i$-th line.
    The updated values are in bold. $C=$
{\footnotesize
    $$\begin{bmatrix}
        -8 & 6 \\
        -7 & 5 \\
        -6 & 4 \\
        -5 & 3 \\
        -3 & 2 \\
        -2 & 1
    \end{bmatrix} 
    \xrightarrow[L1]{}
    \begin{bmatrix}
        \textbf{-7} & \textbf{6} \\
        -7 & 5 \\
        -6 & 4 \\
        -5 & 3 \\
        -3 & 2 \\
        -2 & 1
    \end{bmatrix} 
    \xrightarrow[L2]{}
    \begin{bmatrix}
        \textbf{1} & \textbf{6} \\
        -7 & 5 \\
        -6 & 4 \\
        -5 & 3 \\
        -3 & 2 \\
        -2 & 1
    \end{bmatrix} 
    \xrightarrow[L3]{}
    \begin{bmatrix}
        \textbf{1} & \textbf{6} \\
        -7 & 5 \\
        \textbf{-5} & \textbf{4} \\
        -5 & 3 \\
        \textbf{-2} & \textbf{2} \\
        -2 & 1
    \end{bmatrix} 
    \xrightarrow[L4]{}
    \begin{bmatrix}
        \textbf{1} & \textbf{6} \\
        \textbf{0} & \textbf{5} \\
        \textbf{0} & \textbf{4} \\
        \textbf{0} & \textbf{3} \\
        \textbf{0} & \textbf{2} \\
        \textbf{0} & \textbf{1}
    \end{bmatrix} 
    $$}
\end{example}

Listing~\ref{lst:ebnfJSONaluAdd} specifies the operator $\AddOp$, which requires that both matrices have the same dimensions.

\begin{lstlisting}[caption={Adding two matrices using ALU (EBNF)}, label={lst:ebnfJSONaluAdd}]
alu_add ::= '(*@\textbf{[}@*)', '(*@\textbf{"ADD\_ACC"}@*)', '(*@\textbf{,}@*)',
    '(*@\textbf{[}@*)', string, '(*@\textbf{,}@*)', string, '(*@\textbf{]}@*)', '(*@\textbf{]}@*)';
\end{lstlisting}

Listing \ref{lst:ebnfJSONaluAddEx} reuses the matrices from Listing \ref{lst:ebnfJSONmatEx} to perform the addition of matrices $A$ and $B$ ($\AddOp(A, B)$). 
Example \ref{ex:ebnfJSONaluAddEx} illustrates the operations on two $2 \times 2$ matrices.

\begin{lstlisting}[caption={Example for Adding two matrices using ALU}, label={lst:ebnfJSONaluAddEx}]
["ADD_ACC", ["A","B"]
\end{lstlisting}

\begin{example}[Adding two matrices using ALU] \label{ex:ebnfJSONaluAddEx}
    Let $A$ and $B$ be two matrices to add.
    Let $C$ be the accumulator matrix.
    The updated values are in bold.
    $$
    A = 
    \begin{bmatrix}
        1 & 2 \\
        3 & 4 
    \end{bmatrix}
    , B =
    \begin{bmatrix}
        1 & -2 \\
        2 & -1
    \end{bmatrix}
    \xrightarrow[]{}
    A=
    \begin{bmatrix}
        \textbf{2} & \textbf{0} \\
        \textbf{5} & \textbf{3}
    \end{bmatrix}
    $$
\end{example}

Listing \ref{lst:ebnfJSONalu} defines the top-level ALU declaration. 
This requires a \lstinline|string| field to specify the target matrix for the ALU operations, which must always be the output matrix.

\begin{lstlisting}[caption={ALU Operations (EBNF)}, label={lst:ebnfJSONalu}]
alu_field ::= alu_add | alu_list;
alu ::= '(*@\textbf{"ALU"}@*)', '(*@\textbf{: }@*)', '(*@\textbf{\{}@*)', string, '(*@\textbf{: }@*)',
    '(*@\textbf{[}@*)', alu_field, '(*@\textbf{]}@*)','(*@\textbf{\}}@*)';
\end{lstlisting}

Listing \ref{lst:ebnfJSONaluEx} illustrates a top-level ALU declaration, in which a list of ALU operations is applied to matrix $C$.

\begin{lstlisting}[caption={Example for ALU Operations}, label={lst:ebnfJSONaluEx}]
"ALU": {
    "C": [
        ["MAX", [[0,0], [1,0], 1]],
        ["MAX_IMM", [[0,0], 1, 1]]
    ]
}
\end{lstlisting}

\emph{Store} can write either the entire output matrix or part of it to DRAM memory. 

\begin{lstlisting}[caption={Store Operations (EBNF)}, label={lst:ebnfJSONstore}]
store ::= '(*@\textbf{"STORE"}@*)', '(*@\textbf{: }@*)',  '(*@\textbf{\{}@*)', string, '(*@\textbf{: }@*)', 
    '(*@\textbf{[}@*)', (data_list | string), '(*@\textbf{]}@*)', '(*@\textbf{\}}@*)';
\end{lstlisting}

Listing \ref{lst:ebnfJSONstoreExFull} provides an example of storing the entire matrix $C$ (corresponding to $\StoreOp \big( (C(i))_{i<16} \big)$). 

\begin{lstlisting}[caption={Example for Store Operations (matrix)}, label={lst:ebnfJSONstoreExFull}]
"STORE" : {"C": ["C"]}
\end{lstlisting}

Listing \ref{lst:ebnfJSONstoreEx} shows how to store sub vectors
(corresponding to $\StoreOp \big( (C(0), C(1), C(3), C(5)) \big)$).

\begin{lstlisting}[caption={Example for Store Operations (vectors)}, label={lst:ebnfJSONstoreEx}]
"STORE" : {"C": [
    [[0, 1], 2],
    [[3, 2], 2]
]}
\end{lstlisting}

Listing \ref{lst:ebnfOtherFields} specifies the \lstinline|"NAME"| field used as a suffix for the output binary filenames.
By default, the compiler generates multiple files including \lstinline|instructions.bin| and \lstinline|weight.bin|.
If, for instance, \lstinline|"NAME": "_L1"| the files names become \lstinline|instructions_L1.bin| and \lstinline|weight_L1.bin|.
Finally, it defines the \lstinline|"STRATEGY"| parameter, allowing the selection of one of the four matrix partitioning strategies detailed in Section \ref{sec:extension}.

\begin{lstlisting}[caption={Top-level fields (EBNF)}, label={lst:ebnfOtherFields}]
name ::= '(*@\textbf{"NAME": }@*)', string;
strategy ::= '(*@\textbf{"STRATEGY": }@*)', [1-4];
\end{lstlisting}

As specified in Listing~\ref{lst:ebnfJSON}, the top-level JSON file is structured sequentially. 
It begins with a \lstinline|"NAME"| field, followed by the definitions for the matrices and the subsequent operations: 
a \lstinline|"LOAD"|;
a \lstinline|"GEMM"| or an \lstinline|"ALU"| or both;
and a \lstinline|"STORE"|.
Finally, it ends with an optional \lstinline|"STRATEGY"|.

\begin{lstlisting}[caption={Top-level JSON (EBNF)}, label={lst:ebnfJSON}]
json ::=  '(*@\textbf{\{}@*)', 
    name, '(*@\textbf{,}@*)', 
    matrices, '(*@\textbf{,}@*)',
    load, '(*@\textbf{,}@*)',
    (gemm, ('(*@\textbf{,}@*)', alu)? | alu), '(*@\textbf{,}@*)', 
    store, 
    ('(*@\textbf{,}@*)', strategy)?,
    '(*@\textbf{\}}@*)';
\end{lstlisting}

Listing \ref{lst:JSONexL3} shows the JSON file for the third layer of LeNet-5 (convolution followed by a ReLU activation).

\begin{lstlisting}[caption={Example for the third layer of LeNet-5}, label={lst:JSONexL3}]
{
    "NAME": "_L3",
    "MATRICES": {
        "INPUT": [1,400, "input"],
        "WEIGHT": [400,120, "./wgt_L3.bin"],
        "OUTPUT": [1,120, "output"]
    },
    "LOAD": {
        "INP": ["INPUT"],
        "WGT": ["WEIGHT"]
    },
    "GEMM": ["OUTPUT","INPUT","WEIGHT"],
    "ALU": {
        "OUTPUT": [
            ["MAX_IMM", [[0,1], 0, 120]
        ]
    },
    "STORE": {"OUTPUT": ["OUTPUT"]},
    "STRATEGY": 1
}
\end{lstlisting}

\section{Automated compilation of CNNs} \label{sec:extension}
Prior to this work, the VTA toolchain supported
only single VTA IR compilation (i.e., a unique layer, the associated matrices of which filled within the VTA SRAM).
To deploy full networks, the proposed extended approach consists of:
1) generating multiple VTA IRs (one per layer),
2) generating the CPU code for chaining the layers,
3) generating the VTA data and operations for all layers
and pre-allocating those binaries in the DRAM.
Let us detail these different steps.

\begin{figure}[htb]
    \centering
    \includegraphics[width=0.7\linewidth]{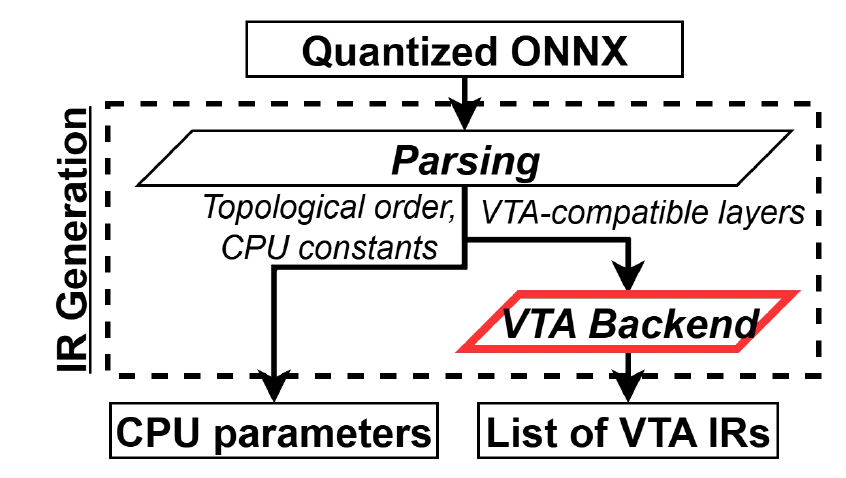}
    \caption{IR Generation/ first compilation stage}
    \label{fig:nn_compiler}
\end{figure}

\textbf{IR Generation.}
The purpose is to translate an ONNX model
into a list of VTA IRs (Figure \ref{fig:nn_compiler}).
For that, the parsing extracts key information from the ONNX model, in particular the topological order, layer-specific attributes, and trained parameters (weights and biases). 
Then, the VTA Backend
transforms layers from the tensor domain to the matrix domain (using im2row-based transformations) and subsequently generates the corresponding VTA IRs (one per layer).

\textbf{CPU code.}
The CPU is in charge of executing operations that cannot be offloaded to the VTA,  orchestrating the VTA offloading (which layers / parts of layers are executed and in which order),
and rearranging matrices computed by the VTA for the next layers.
Some memory reshaping between layers is needed because the outputs computed by the VTA are often tensors while the next offloaded layers expect im2row matrix format.
The \emph{chaining} performed by the CPU entails both managing the data flow and enforcing
memory organisation consistency between layers. 
Practically, 
the CPU code is hard coded
and takes as input some parameters depending on the CNN.
These \emph{CPU parameters} are computed by the compiler (see Figure \ref{fig:nn_compiler})
and are stored in a text file which contains a set of constants.

\begin{figure}[hbt]
    \centering
    \includegraphics[width=\linewidth]{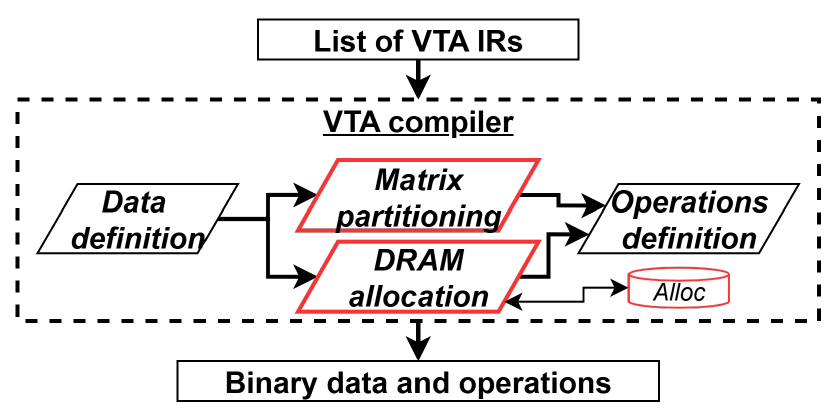}
    \caption{Enhanced VTA compiler}
    \label{fig:enhanced_vta_compiler}
\end{figure}

\textbf{Data and Instructions Generation.}
The chosen approach to execute multiple VTA IRs consists in storing statically
all data and operations in the DRAM.
For that, we update the DRAM allocation module to 
allocate a dedicated address space for each layer
(see Figure \ref{fig:enhanced_vta_compiler}).

\begin{figure}[hbt]
    \centering
    \includegraphics[width=0.8\linewidth]{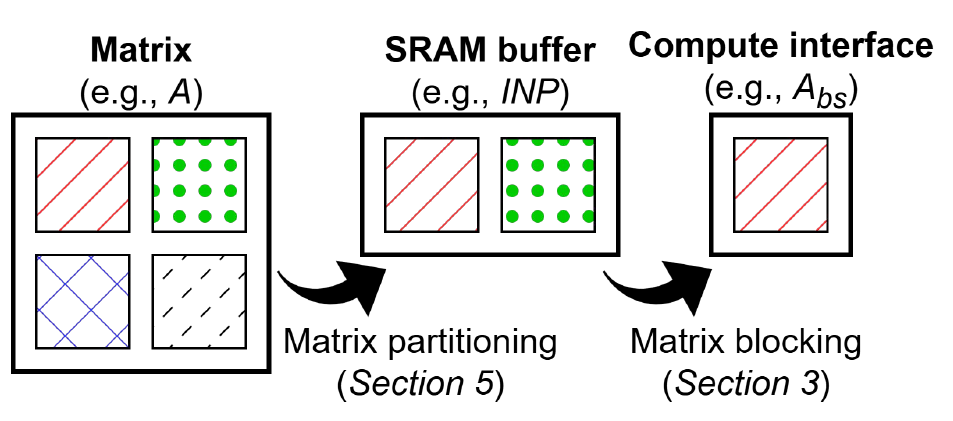}
    \caption{Two levels of matrix decomposition: the matrix comprises four blocks, the SRAM capacity is limited to two. Atomic operations process a single block.}
    \label{fig:blocking_vs_partitioning}
\end{figure}

Executing a matrix operation on the VTA requires two levels of matrix decomposition, as illustrated in Figure \ref{fig:blocking_vs_partitioning}:
1) Matrix Partitioning: due to limited SRAM capacity, the compiler may have to partition large matrices operations into multiple smaller partial operations.
2) Matrix Blocking: the VTA enforces specific data shape requirements (cf. Section \ref{sec:background}), which necessitates block-based matrix operations (cf. Section \ref{sec:formal_op}).

Our prior work supported only matrix blocking. 
We extend the VTA compiler with a Matrix Partitioning module
that automatically detects whether an input, weight, or output matrix exceeds the VTA's SRAM capacity.
It decomposes the operations and data
into a set of matrix-block operations that fit within the SRAM.
There are multiple strategies to partition the matrix
and Section \ref{sec:strategy} details the constraints that a valid strategy must fulfill.
Currently, the matrix partitioning module implements four heuristic strategies
(denoted $1, 2, 3, 4$) for GEMM and a unique one for ALU.
By default, the IR generation selects for GEMM the Strategy 1, but the VTA IR can be modified
to select any of them.

\begin{figure}[htb]
    \centering
    \includegraphics[width=0.8\linewidth]{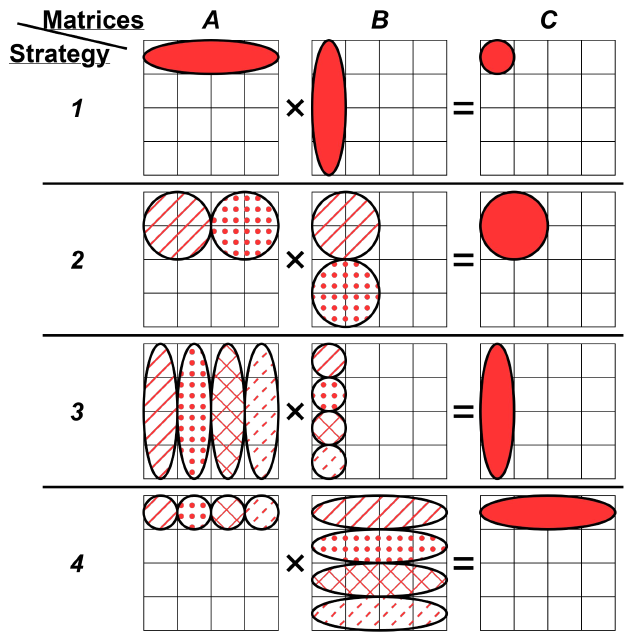}
    \caption{The four implemented matrix partitioning strategies}
    \label{fig:strategy}
\end{figure}

Figure \ref{fig:strategy} illustrates those 4 strategies on an example
where the buffer capacity is limited to 4 blocks. 
The oval patterns identify the specific blocks involved in the computation.
Blocks in matrices $\matA$ and $\matB$ marked with matching symbols are multiplied, and their products are accumulated. 
The final result is represented by the solid oval in matrix $\matC$.
Strategy 1 computes a single block in $\matC$ by multiplying a row of $\matA$ with a column of $\matB$.
Strategy 2 performs square block-based computation to produce a corresponding square sub-matrix in $\matC$.
Strategies 3 and 4 are symmetrical: 
Strategy 3 multiplies a column of $\matA$ by a single block of $\matB$ to generate a column in $\matC$, 
whereas Strategy 4 multiplies a single block of $\matA$ by a row of $\matB$ to generate a row in $\matC$.
In this example, computing four blocks of $\matC$ requires four partitioning steps using Strategies 1, 3, and 4, whereas Strategy 2 requires only two.

\section{Matrix partitioning strategies}
\label{sec:strategy}
The input of the matrix partitioning
module is a VTA IR such as the one of the LeNet-5 shown in listing
\ref{lst:JSONexL3}.
The generic produced IR for any CNN layer
follows the same
approach as shown in listing \ref{lst:gen-IR}.

\begin{lstlisting}[caption={Generic VTA IR}, label={lst:gen-IR}]
  "MATRICES": {
    "A": [alpha, lambda, "input"],
    "B": [lambda, beta, "./wgt.bin"],
    "X": [alpha, beta, "./acc.bin"],
    "C": [alpha, beta, "output"]
    },
  "LOAD": {
    "INP": ["A"],
    "WGT": ["B"],
    "ACC": ["X"],
    },
  "GEMM": ["C","A","B"],
  "ALU": {
    "OUTPUT": [
      OP1,
      ...,
      OPk
     ]
    },
    "STORE": {"C": ...}
\end{lstlisting}

The operations of a VTA IR involve
(at most) 4 block-matrices: $\matA$, $\matB$, $\matX$, and $\matC$ of size
$\alpha \times \lambda$,
$\lambda \times \beta$,
$\alpha \times \beta$
and $\alpha \times \beta$ respectively.

\begin{definition}[Condition to trigger a partitioning]
  \label{def:overflow}
  A partitioning is needed if one of the matrices does not fit into its respective buffer (i.e., memory overflow):
  $\alpha \times \lambda > \memInp$ or $\lambda \times \beta > \memWgt$
  or $(\alpha \times \beta) \times \blockSize > \memAcc$.
\end{definition}

The VTA IR involves multiple matrix operations:
at most one GEMM operation,
k ALU operations (with $k\geq 0$).
If the matrix partitioning module is triggered, at least one
of those operations has to be decomposed.
A matrix operation
that does not fit into the VTA SRAM
is split
into a set of sub-matrix operations
such that each one fits into the SRAM.
Then these  sub-matrix operations are implemented
as multiple successive offloadings onto the accelerator.
There are several valid ways
of decomposing the initial matrix operation but
some constraints
have to be respected.

\subsection{Decomposition of a \BGemmOp}
The operation
$\matC := \matC + \matA \times \matB$
consists in executing all block operations
as shown in definition \ref{def:block_matmul},
i.e.
$ \BGemmOp(\matC, \matA, \matB) = \{ \GemmOp(\matBlock{\matC}{i \times \beta + j}, \matBlock{\matA}{i \times \lambda + k}, \matBlock{\matB}{k \times \beta + j}) \allowbreak | i \in [\![0, \alpha -1]\!], j < [\![0, \beta -1]\!], k \in [\![0, \lambda -1]\!]
\}$.
Any execution order of the block operations is valid.

\begin{property}[Rewriting a \BGemmOp]
Let us consider the set of triplets
$P_{(\matC, \matA, \matB)} = \{(l,p,m) |
l=i \times \beta + j, p = i \times \lambda + k, m = k \times \beta + j,
i \in [\![0, \alpha -1]\!], j < [\![0, \beta -1]\!], k \in [\![0, \lambda -1]\!]
\}$.
Then,
$ \BGemmOp(\matC, \matA, \matB) = \cup_{(l,p,m) \in P_{(\matC, \matA, \matB)}}
\{ \GemmOp(\matBlock{\matC}{l}, \matBlock{\matA}{m}, \matBlock{\matB}{p})\}$.
\end{property}

\begin{definition}[Valid decomposition of $ \BGemmOp(\matC, \matA, \matB)$]
Let $l \leq |P_{(\matC, \matA, \matB)}|$,
  a valid decomposition of  $ \BGemmOp(\matC, \matA, \matB)$ into $l$ offloadings
  is a partition $\Pi = \{P_1, \ldots, P_l\}$
  of $P_{(\matC, \matA, \matB)}$
  such that
  $\cup_i P_i = P_{(\matC, \matA, \matB)}$,
  $\forall i\not= j, P_i \cap P_j = \emptyset$,
  $|P_i|\leq \textit{min}(\memInp,\memWgt, \memAcc)$.
  The last constraint states that the blocks of $\matA$ (resp. $\matB$, $\matC$) of any element $P_i$
  fits in $\INP$ (resp. $\WGT$, $\ACC$).
\end{definition}

\begin{example}[Strategy 1 of Figure \ref{fig:strategy}]
  \label{ex-indices}
Looking at strategy 1,
$P_1$ in red is defined by
$\{(0,0,0),(0,1,4), (0,2,8)$, $(0,3,12)\}$.
The strategy consists in computing
the block $\matBlock{\matC}{i}$
for $i=0$ to $\alpha\times\beta$ with 
partition $P_i$.
Thus, $P_2=\{(1,0,1),(1,1,5), (1,2,9), (1,3,13)\}$.
\end{example}

\begin{example}[Strategy 2 of Figure \ref{fig:strategy}]
  For strategy 2,
  $P_1$ corresponds to the dashed group;
  thus $P_1$ $=$ $\{
    (0,0,0) \comma (0,1,4) \comma (1,0,1) \comma (1,1,5) \comma
    (4,4,0) \comma (4,5,4) \comma (5,4,1) \comma (5,5,5)
  \}$
  and $P_2$ the dotted pattern group, i.e.,
  $P_2$ $=$ $\{
    (0,2,8) \comma (0,3,12) \comma (1,2,9) \comma (1,3,13) \comma
    (4,6,8) \comma (4,7,12) \comma (5,6,9) \comma (5,7,13)
  \}$.
  The strategy consists in computing
the a block of $\matC$ with 2 partitions.
\end{example}

\begin{example}[Strategy 1 with memory overflow]
  If a line of $\matA$ does not fit in $\INP$, the strategy 1
  computes $\matBlock{\matC}{i}$ with multiple $P_{i_k}$.
Let us suppose for the example that only 2 blocks of $\matA$ and  $\matB$
can fit into the SRAM.
Then,   $\matBlock{\matC}{0}$,
is computed with 2 partitions:
$\{(0,0,0) ,(0,1,4)\}$
and $\{(0,2,8)$, $(0,3,12)\}$.
\end{example}

\subsection{Decomposition of a \textit{bALU}}
The definition \ref{def:alu_vector-vector} details that
  $\BAluOp{\OpElem}(\matX, \matY)$
consists of a set of $\AluOp{}$
$
        \{ 
            \AluOp{\OpElem}(\matBlock{\matX}{i \times \beta}, \matBlock{\matY}{i \times \beta})
            | i \in [\![ 0, \beta -1 ]\!] 
        \}
        $.
        Similarly to \BGemmOp, we rewrite $\BAluOp{\OpElem}$
        to refer to the indices involved in the ALU block.

\begin{property}[Rewriting a $\BAluOp{\OpElem}$]
Let us consider the set of pairs
$P_{(\matX, \matY)} = \{(l,p) |
l=i \times \beta, p = i \times \beta,
i \in [\![0, \beta -1]\!]\}$ and
  $Q_{\matX,c}=\{l | l=i \times \beta, i \in [\![0, \beta -1]\!]\}$.
Then,
$ \BAluOp{\OpElem}(\matX, \matY) =
\cup_{(l,p) \in P_{(\matX, \matY)}}
\{ \AluOp{\OpElem}(\matBlock{\matX}{l}, \matBlock{\matY}{p})\}$
and
$ \BAluOp{\OpElem}(\matX, c) =
\cup_{(l) \in Q_{(\matX, c)}}
\{ \AluOp{\OpElem}(\matBlock{\matX}{l}, c)\}$.
\end{property}

\begin{figure}[htb]
    \centering
    \includegraphics[width=.9\linewidth]{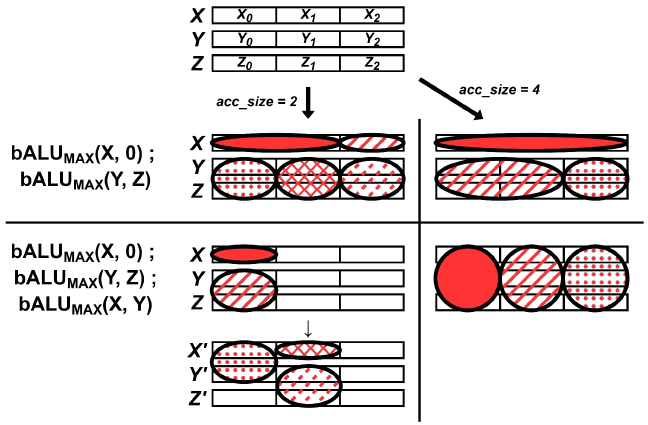}
    \caption{Decomposing a succession of $\BAluOp{\OpElem}$}
    \label{fig:alu-strategy}
\end{figure}

We have implemented a unique strategy to handle large ALU operations as shown in Figure \ref{fig:alu-strategy}.
An immediate operation
$\AluOp{\OpElem}(\matBlock{\matX}{l}, c)$ such that $\forall p, (l,p) \not\in P_{(\matX, \matY)}$ (the vector $\matBlock{\matX}{l}$ is never reused), the operation for that vector is done through the line as shown (see top
of the figure for 2 different $\memAcc$ size).
Otherwise, execution proceeds column-by-column (see bottom of the figure) 
and batches as many columns as the $\ACC$ buffer permits.

\subsection{Decomposition of a VTA IR}
A VTA IR contains at most one GEMM and
we have seen that the block GEMM can be executed in any order.
A $\BAluOp{\OpElem}(\matX, \matY)$
is also decomposed into parts of the vectors and there is no constraint on the order of the $\AluOp{\OpElem}$.
However, there are some constraints on the order
between:
\begin{itemize}
  \item a
$\BAluOp{\OpElem}(\matX, \matY)$ following a GEMM if $\matX$
    or $\matY$ is a line of the matrix $\matC$
  \item two successive $\BAluOp{\OpElem}(\matX, \matY)$ and
    $\BAluOp{\OpElem}(\matX', \matY')$ if $\matX=\matX'$ or  $\matX=\matY'$.
\end{itemize}

A strategy
consists in taking a partition of  $ \BGemmOp$, then possibly of partition
of  $\BAluOp{\OpElem}(\matX, \matY)$ and load the adequate buffers.
Our strategy is to execute first the $ \BGemmOp$,
then the $\BAluOp{\OpElem}(\matX, \matY)$.

\begin{figure}[htb]
    \centering
    \includegraphics[width=.9\linewidth]{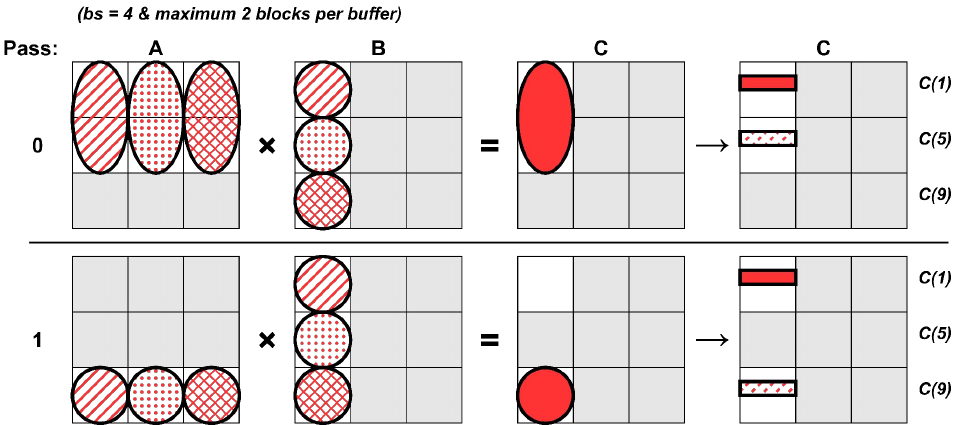}
    \caption{Possible combination between GeMM and ALU matrix partitioning}
    \label{fig:gemm_alu_strategy}
\end{figure}

Future work consists of formally defining and computing
optimal offloading strategy to minimise the number of loads as proposed in \cite{BenjaminH}. 
Figure \ref{fig:gemm_alu_strategy} illustrates a possible GeMM and ALU partitioning strategy, applied to a sequence where $\BGemmOp(\matC, \matA, \matB)$ is followed by $\big( \BAluOp{\OpElem}(\matC(1), \matC(5)) \comma \BAluOp{\OpElem}(\matC(1), \matC(9)) \big)$.
This method extends GEMM Strategy 3 by interleaving GEMM and
column-wise ALU operations.

\section{Experiments: YOLO-NAS} \label{sec:experiments}
The compilation is evaluated against three objectives:
1) Ensure CNN compilation and correctness by comparing results to a reference function;
2) Measure memory overhead relatively to the original ONNX model;
3) Analyse matrix partitioning impact on the number of instructions.

The evaluation is done with a quantized \emph{YOLO-NAS} (You Only Look Once - Neural Architecture Search) CNN model \cite{yolo_all}. 
This CNN is representative of industrial-grade models because: 
it contains main standard operators (QuantizeLinear, QLinearConv, QLinearMul, QLinearAdd, QLinearConcat, MaxPool, ConvTranspose, DequantizeLinear), 
features parallel computation branches,
and is used in industrial applications \cite{yolo_ex1,yolo_ex12}. 
Furthermore, YOLO-NAS contains large tensors that exceed the VTA SRAM capacity, thereby triggering matrix partitioning.

\textbf{Parsing and Compilation.}
The quantized YOLO-NAS comprises 146 operators,
among which 108 operators, including QLinearConv, QLinearMul, and MaxPool, are VTA-compatible. 
The remaining 38 operators (e.g., QuantizeLinear, QLinearAdd, QLinearConcat, ConvTranspose, DequantizeLinear) are executed on the CPU, as they require floating-point operations.

The parsing and VTA IR generation phases take a few seconds. 
Compiling the 108 VTA IRs with Strategy 4 takes approximately 2 minutes and generates 10.8 million instructions and 9.1 million UOPs.

\begin{figure}[htb]
    \centering
    \includegraphics[width=\linewidth]{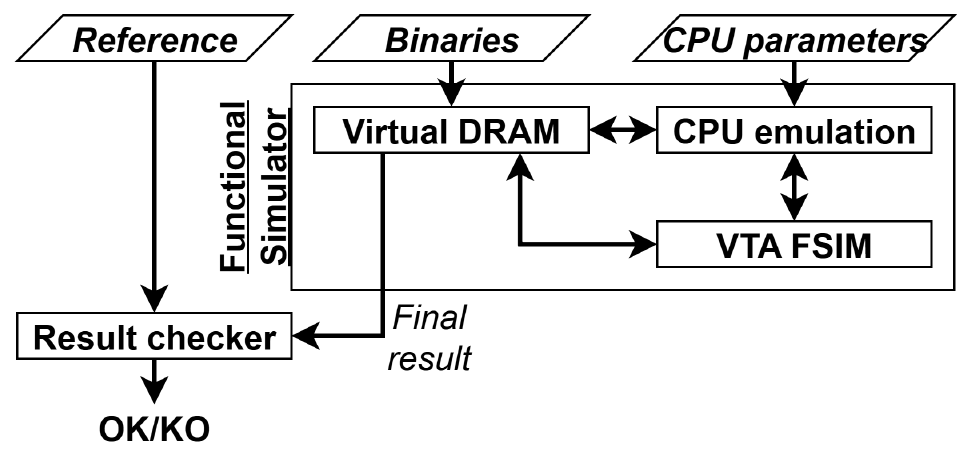}
    \caption{Functional simulator}
    \label{fig:fsim}
\end{figure}

\textbf{Simulation.}
We simulate with a functional C++ simulator
the execution of the CNN on the processor composed of a
CPU and a VTA accelerator.
The simulator, see Figure \ref{fig:fsim},
incorporates the original virtual DRAM and VTA functional simulation modules \cite{VTA_paper}. 
We add a CPU emulation module configured 
using the CPU parameters generated during compilation. 
This module orchestrates the VTA execution and performs layer chaining and data reshaping.
The simulation of the YOLO-NAS execution takes approximately 3 minutes, most of the time is spent
by the CPU manipulating data between layers. 

\textbf{Correctness.}
Correctness is evaluated via bit-wise comparison of the results. 
The tests are conducted with randomly generated inputs spanning the entire int8 range $[\![-128, 127]\!]$.
First, the reference output was generated using ONNX Runtime. 
Across ten executions, the majority of differences are $\pm 1$, affecting up to 35\% of the data. 

The discrepancies originate from the QLinearConv implementation. 
Specifically, ONNX Runtime applies optimisations to this operator that deviate from the ONNX mathematical description. 
Consequently, we implemented a reference function in Numpy that adheres to the mathematical definition. 
The simulation results are bit-accurate when compared to this Numpy reference across the ten executions.

\textbf{Memory overhead.}
Table \ref{tab:yolo_compilation} compares the memory footprint of the original ONNX model and the compiled output. 
The graph encompasses the network structure, topological order, tensor shapes, and rescaling factors. 
The weights and the biases correspond to the trained parameters.
Finally, the instructions encompass the binary instructions and UOPs.

\begin{figure}[htb]
    \centering
    \includegraphics[width=0.7\linewidth]{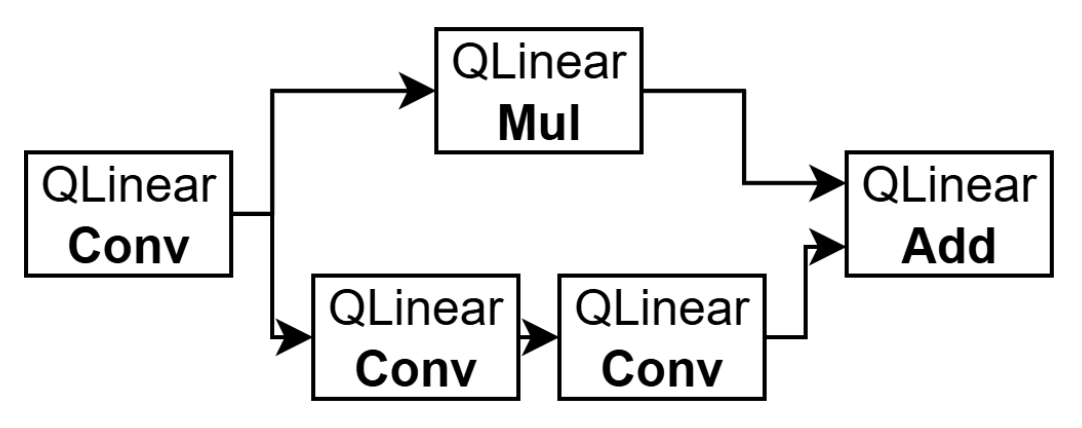}
    \caption{Recurring YOLO-NAS pattern}
    \label{fig:pattern}
\end{figure}

For comparison purposes, three ONNX models were compiled:
1) A single QLinearConv layer with an input dimension of $1 \times 3 \times 1024 \times 1024$ and an output dimension of $1 \times 32 \times 512 \times 512$ (both in NCHW format), including bias;
2) The recurring pattern of the YOLO-NAS architecture, as shown in Figure \ref{fig:pattern};
3) The full quantized YOLO-NAS model.

\begin{table}[hbt]
        \centering
        \begin{tabular}{|S{c}|S{c}|S{c}|S{c}|} 
            \cline{2-4}
            \multicolumn{1}{c|}{} & \multicolumn{3}{c|}{\textbf{Memory size}} \\
            \cline{2-4}
            \multicolumn{1}{c|}{} & QLinear & Pattern & YOLO \\
            \multicolumn{1}{c|}{} & Conv &  & NAS \\
           \hline
           ONNX Graph & 912 B & 4,935 B & 122 KiB \\
           \hline
           Compiled & 364 B & 1,421 B & 39 KiB \\
           Graph & (-60.1\%) & (-71.2\%) & (-68.0\%) \\
           \hline
           ONNX Weights & 864 B & 20,480 B & 5,597 KiB \\
           \hline
           Compiled & 1,024 B & 20,736 B & 5,643 KiB \\
           Weights & (+18.5\%) & (+1.3\%) & (+0.8\%) \\
           \hline
           ONNX Biases & 128 B & 384 B & 33 KiB \\
           \hline
           Biases & 32 MiB & 32 MiB & 297 MiB \\
           \hline
           Instructions & 3,073 KiB & 12 MiB & 93 MiB \\
           \hline
        \end{tabular}
    \caption{Memory overhead between ONNX and compiled CNN}
    \label{tab:yolo_compilation}
\end{table}

The compiled graph is more compact than the ONNX graph, as it retains only matrix information rather than the more expensive tensor representation.
In contrast, the compiled weights are slightly larger due to the padding required to form square blocks. 
However, the most significant impact on memory footprint stems from the biases. 
While ONNX stores biases as compact vectors, the compiler expands these into matrices to initialise the VTA's accumulator buffer, resulting in heavy data replication. 
For instance, the $1 \times 32$ bias vector from the QLinearConv operator is transformed into a $262,144 \times 32$ matrix (plus padding). 
Finally, the compilation footprint includes generated VTA instructions and UOPs.

\textbf{Matrix partitioning impact.}
The four matrix partitioning strategies, presented in Section \ref{sec:extension} (Figure \ref{fig:strategy}), are assessed.
The choice of strategy significantly impacts the instruction count, as shown in Table \ref{tab:strat_yolo}. 
However, it does not affect the number of UOPs at all. 
This is because the strategy alters the memory access patterns (load/store) and execution order, but the underlying set of atomic operations remains constant (executed exactly once per operation).

\begin{table}[!ht] 
        \centering
        \begin{tabular}{|S{c}|S{c}|S{c}|} 
            \hline
            \textbf{Strategy} & \textbf{Number of} & \textbf{Number of} \\
            \textbf{} & \textbf{instructions} & \textbf{UOPs} \\
            \hline
            \textbf{1} & 7,505,759 & 9,125,292 \\
            \hline
            \textbf{2} & 7,246,143 & 9,125,292 \\
            \hline
            \textbf{3} & 6,348,149 & 9,125,292 \\
            \hline
            \textbf{4} & 10,765,279 & 9,125,292 \\
            \hline
        \end{tabular}
    \caption{Strategy choice impacts on the number of instructions for YOLO-NAS}
    \label{tab:strat_yolo}
\end{table}

Table \ref{tab:strat_shape} illustrates the impact of matrix shape on the number of instructions required by each strategy to perform a GEMM operation. 
We define matrices $\matA$, $\matB$, and $\matC$ with dimensions $\alpha \times \lambda$, $\lambda \times \beta$, and $\alpha \times \beta$. 
In each case, the total computational workload is fixed at $\alpha \times \lambda \times \beta = 2^{28}$ operations, resulting in 65,521 UOPs. 
Percentage variations are calculated relative to Strategy 1.
The observations demonstrate that strategy selection is highly shape-dependent: the optimal choice in terms of instruction count for one configuration can be the worst for another (e.g., Strategy 4 in Case 1 versus Case 2). 
Performance for Strategy 1 depends on the total number of output elements, with smaller output sizes yielding better results.
Strategies 3 and 4 show symmetrical behavior: Strategy 3 is more efficient with fewer output columns, while Strategy 4 performs best with fewer output rows. 
Finally, Strategy 2 offers a good compromise, as it is consistently reliable and never the worst-performing option in terms of instruction count.

\begin{table}[!ht]
    \centering
    \begin{tabular}{|S{c}|S{c}|S{c}|} 
        \hline
        \textbf{Shape} & \textbf{Strategy} & \textbf{Number of} \\
        \textbf{} & \textbf{} & \textbf{instructions} \\
        \hline
        \textbf{CASE 1} & 1 & 49,157 \\ 
        \cline{2-3}
        $A = 2^{5} \times 2^{7}$ & 2 & 49,925 (+1.6\%) \\ 
        \cline{2-3}
        $B = 2^{7} \times 2^{16}$ & 3 & \textit{143,365} (+191.6\%) \\ 
        \cline{2-3}
        $C = 2^{5} \times 2^{16}$ & 4 & \textbf{10,309} (-79.0\%) \\ 
        \hline

        \textbf{CASE 2} & 1 & 49,157 \\ 
        \cline{2-3}
        $A = 2^{16} \times 2^{7}$ & 2 & 10,757 (-78.1\%) \\ 
        \cline{2-3}
        $B = 2^{7} \times 2^{5}$ & 3 & \textbf{10,309} (-79.0\%) \\ 
        \cline{2-3}
        $C = 2^{16} \times 2^{5}$ & 4 & \textit{143,365} (+191.6\%) \\ 
        \hline
%
        
        \textbf{CASE 3} & 1 & \textbf{2,181} \\ 
        \cline{2-3}
        $A = 2^{7} \times 2^{14}$ & 2 & 8,454 (+287.6\%) \\ 
        \cline{2-3}
        $B = 2^{14} \times 2^{7}$ & 3 & \textit{32,845} (+1,406.0\%) \\ 
        \cline{2-3}
        $C = 2^{7} \times 2^{7}$ & 4 & \textit{32,845} (+1,406.0\%) \\ 
        \hline
        
    \end{tabular}
    \caption{GEMM shape impacts on the number of instructions, bold numbers denote the best results, while italicised ones denote the worst}
    \label{tab:strat_shape}
\end{table}

\textbf{Discussion and limitations.} 
The compilation chain is capable of automatically compiling and simulating full quantized ONNX models on the VTA. 
The generated models achieve bit-accurate consistency with the ONNX mathematical reference, despite minor deviations from ONNX Runtime optimisations.
Furthermore, the total pipeline execution time remains acceptable, requiring only a few minutes.

Nevertheless, several limitations remain. 
First, the compilation relies heavily on the CPU due to floating-point operations within the operators (e.g., rescaling). 
Adopting a fixed-point real representation
would enable the offloading of additional operators (such as QLinearAdd and ConvTranspose) and support hardware-based post-operation rescaling.
Additionally, investigating methods to minimise data reorganisation between layers could significantly reduce the CPU workload.

Second, the compiled models exhibit a significant memory overhead due to the biases. 
This could be mitigated by replicating bias vectors at runtime (e.g., using load or ALU operations) rather than expanding them during compilation. 

Third, the assessment of matrix partitioning relies solely on instruction count. 
However, instruction count does not directly correlate with VTA latency. 
To optimise the selection of a matrix partitioning strategy (and identify the most efficient approach), a cycle-accurate simulation is required.
While a cycle-accurate CHISEL simulator exists, it is currently limited to the Compute module. 
Extending this simulator to encompass the entire VTA, including the Load and Store modules, would yield valuable insights into execution timing and identify performance bottlenecks.

\section{Conclusion} \label{sec:conclusion}
The extended VTA stand-alone compiler compiles industrial-grade CNN models from int8 quantized ONNX files into VTA binaries and CPU parameters.
A functional C++ simulator enables the execution of the compiled model and verifies the results against either the ONNX Runtime or a Numpy reference implementation.

This work represents a significant step toward establishing the VTA as a viable open-source matrix accelerator.
However, three limitations have
been identified and will be addressed in future work.
In parallel, we will also port the VTA to the UltraScale+ and
evaluate the real behaviour of compiled models on the target.


\section*{Acknowledgment}
This work has benefited from the AI cluster ANITI2 funded by the French government through the ANR under the France 2030 program (grant ANR-23-IACL-0002).

\bibliographystyle{abbrv}
\PHMbibliography{erts}

\end{document}